\documentclass{aa}

\usepackage[varg]{txfonts}
\usepackage{siunitx}
\DeclareSIUnit\um{\micro\meter}
\DeclareSIUnit\Msun{M_\odot}

\usepackage{url}

\usepackage[tbtags,fleqn]{amsmath}
\usepackage{amsfonts}

\usepackage[x11names]{xcolor}
\usepackage{graphicx}
\usepackage{natbib}


\begin{document}

\title{Optimal extinction measurements}
\subtitle{I. Single-object extinction inference}
\author{Marco Lombardi}
\mail{marco.lombardi@unimi.it}
\institute{%
  University of Milan, Department of Physics, via Celoria 16, I-20133
  Milan, Italy} 
\date{Received ***date***; Accepted ***date***}

\abstract{%
  In this paper we present XNICER, an optimized multi-band extinction
  technique based on the extreme deconvolution of the intrinsic colors
  of objects observed through a molecular cloud. XNICER follows a
  rigorous statistical approach and provides the full Bayesian
  inference of the extinction for each observed object. Photometric
  errors in both the training control field and in the science field
  are properly taken into account. XNICER improves over the known
  extinction methods and is computationally fast enough to be used on
  large datasets of objects. Our tests and simulations show that this
  method is able to reduce the noise associated with extinction
  measurements by a factor 2 with respect to the previous NICER
  algorithm, and it has no evident bias even at high extinctions.}
\maketitle

\section{Introduction}

Understanding the anatomy of molecular clouds is critical to decipher
the process of star and planet formation. However, since molecular
clouds are mainly composed of cold ($\sim \SI{10}{K}$) molecular
hydrogen and helium, which are virtually invisible, studies of these
objects have to rely on rare tracers and extrapolate their small
abundancies to obtain the full mass distribution.

Of the techniques that are used to study molecular clouds, the ones
based on optical and infrared extinction are probably the most
consistent \citep{2009ApJ...692...91G}. They are based on what is
widely thought to be the most reliable gas tracer at our disposal:
dust. Moreover, they use a very small set of assumptions and are free
from many of the uncertainties that plague other methods. Infrared
extinction, in particular, has been successfully employed to
investigate the interstellar medium at many different scales: from
small cores (e.g., \citealp{2001Natur.409..159A}), to giant molecular
clouds \citep{2010A&A...512A..67L}, to the entire sky (using star
counts, as described by \citealp{2011PASJ...63S...1D} and
\citealp{2013PASJ...65...31D}, or color excess, as described by
\citealp{2016A&A...585A..38J}).

Extinction measurements are not only powerful per se: they are also
often used to calibrate other techniques, so as to reduce many
uncertainties. For example, submillimeter dust emission observations
can be used to produce exquisite dust emission maps, which often
present spectacular views of entire molecular clouds at relatively
high resolution. These maps, however, would be scientifically of
little use without a proper calibration, which is often achieved
using (lower resolution) extinction maps of the same area (e.g.,
\citealp{2014A&A...566A..45L, 2016A&A...587A.106Z}). Similarly,
extinction is used to calibrate maps of near-infrared scattered light
(\citealp{2006A&A...457..877J}; see also, e.g.,
\citealp{2013A&A...558A..44M}). Finally, extinction is crucial also to
infer the X-factor of radio observations of different molecules.

As early recognized by \citet{1994ApJ...429..694L}, extinction is best
measured from the color excess of background stars. Most studies since
then have been carried out in the near-infrared (NIR), because at
these wavelengths molecular clouds are more transparent than at
visible bands, and as a result, one is able to probe denser
regions. An additional important benefit is that most stellar objects
have similar colors in the NIR, which makes the measurement of
the color excess more accurate for these objects. Similarly, different
reddening laws show little scatter in the NIR, and this makes infrared
extinction measurements very consistent \citep{2005ApJ...619..931I,
  2007ApJ...663.1069F, 2013A&A...549A.135A}.

The original NIR extinction studies have been carried out using the
Near Infrared Color Excess (NICE) algorithm
\citep{1994ApJ...429..694L} using only two NIR bands (typically, the
combination $H - K$). Subsequently, the algorithm has been refined
into NICER to take advantage of multi-band photometry and to include a
better description of the errors involved \citep{2001A&A...377.1023L}
while retaining much of its simplicity. NICER has since then been used
in many different studies of molecular clouds (see, e.g.,
\citealp{2006A&A...454..781L, 2011A&A...535A..16L,
  2014A&A...565A..18A, 2016A&A...585A..38J}), in many cases involving
data from the Two Micron All Sky Survey (2MASS,
\citealp{2006AJ....131.1163S}).

The advent of more powerful instruments and deeper observations has
shown some of the limitations of the NICER algorithm, which describes
the intrinsic colors of stars as a multivariate Gaussian
distribution. The reality is much more complex, and this has
stimulated a number of authors to investigate more advanced methods to
meet the challenges posed by the new data \citep{2005A&A...438..169L,
  2016A&A...585A..78J, 2017A&A...601A.137M}. The need for a better
understanding of the whole process has been triggered not only from
NIR, but also from deep optical surveys such as Pan-STARRS
\citep{2010SPIE.7733E..0EK}, and is likely to be even stronger with
the future releases of the \textit{Gaia} mission
\citep{2016A&A...595A...1G}.

This paper follows this trend and aims at providing a more accurate
and reliable framework to perform extinction studies from multi-band
photometry. The method presented here, named XNICER, shares much of
the simplicity and speed of the NICER algorithm while allowing a
complex description of the intrinsic (unextinguished) colors of
background objects. XNICER follows a rigorous statistical approach and
provides a method for obtaining a full Bayesian inference of the
extinction of an object given a training set (control field). In this
first paper we limit our investigation to the simplest version of the
algorithm and to measurements of individual extinctions; more complex
analyses are deferred to follow-up papers.

The paper is structured as follows. In
Sect.~\ref{sec:method-description} we describe XNICER and all
necessary steps and improvements needed to use it best. A comparison
with alternative techniques is given in
Sect.~\ref{sec:comp-with-diff}. We then describe several tests that
were carried out in a control field to assess the merits and
limitations of XNICER (also compared to alternative methods) in
Sect.~\ref{sec:contr-field-analys}. A sample application to Orion~A is
briefly discussed in Sect.~\ref{sec:sample-appl-orion}. Finally, we
provide an overview of the implementation in
Sect.~\ref{sec:implementation} and summarize the results of this paper
in Sect.~\ref{sec:conclusions}.

\section{Method description}
\label{sec:method-description}

\begin{figure}[tb!]
  \centering
  \includegraphics[width=\hsize]{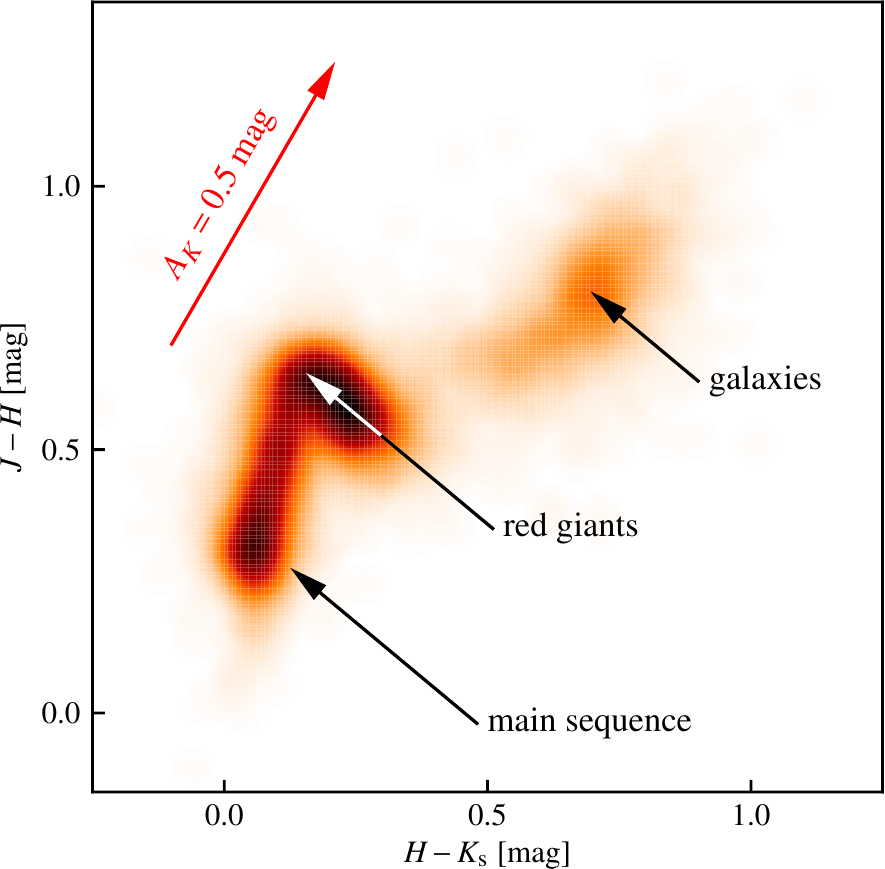}
  \caption{Color distribution of stars with accurate photometry
    (photometric error below \SI{0.1}{mag} in both colors). The
    various clumps have a simple direct interpretation, as indicated
    by the black arrows. The red arrow indicates the shift operated by
    a \SI{0.5}{mag} $K$-band extinction.}
  \label{fig:1}
\end{figure}

XNICER works similarly to other color-excess methods. The method is
purely empirical and requires photometric measurements on a control
field, where the extinction is presumably absent, and on a science
field, containing the area of interest. More precisely, for each
object in both fields, we presume to have at our disposal magnitudes
and associated errors in different bands.

XNICER, in many senses, generalizes the NICER technique
\citep{2001A&A...377.1023L} to situations where the intrinsic color
distribution of background objects cannot be described in terms of
mean and covariance alone. In order to better understand how XNICER
works in a qualitative way, it is useful to consider the distribution
of intrinsic star colors; later on, we will provide a rigorous
statistical description of the method. For this purpose, we make use
of the Vienna Survey in Orion dataset (VISION,
\citealp{2016A&A...587A.153M}), a deep NIR survey of the Orion~A
molecular cloud in the $J$, $H$, and $K_\mathrm{s}$ passbands. The use
of VISION ensures that we test the method on relatively complex data,
including also extended background objects (mostly galaxies) in
addition to point-like ones. We stress, however, that XNICER can be
applied equally well to any number and combination of photometric
measurements, including optical and mid-infrared ones.

Figure~\ref{fig:1} shows the distribution in color space of objects in
the control field with accurate photometric measurements. We refer to
the \textup{intrinsic} distribution of object colors throughout, that
is,\ the distribution that would be observed with measurements that
are unaffected by photometric errors and have no extinction, such as
the intrinsic probability density function (PDF, or iPDF for
short). The data shown in Fig.~\ref{fig:1} follow to a first
approximation the iPDF if we ignore the (relatively small) photometric
errors present in these data. Clearly, this figure shows that the iPDF
is multimodal and cannot be accurately described by a simple Gaussian:
this is, however, implicitly what the NICER technique does.

When a uniform extinction is present, the color distribution of stars
is (to a first approximation, see below
Sect.~\ref{sec:numb-counts-extinct}) shifted along the reddening
vector. This suggests that we can infer the extinction that affects
each object by tracing the observed colors of each object back along
the reddening vector and considering the amplitudes (i.e.,\ the
values) of the iPDF along this line. These can be directly interpreted
as an extinction probability distribution for the star. The same
method was essentially also used by \citet{2017A&A...601A.137M} in the
PNICER technique.

In reality, this simple scheme has some problems that need to be
solved before it can be applied in a rigorous way to real data:
\begin{itemize}
\item any background object will likely have photometric errors that
  will affect the observed colors and that need to be taken into
  account;
\item some objects will also have missing bands, a fact that clearly
  needs to be taken into account;
\item the same two issues mentioned above (photometric errors and
  missing bands) will also affect the objects observed in the control
  field, and this will affect our capability of inferring the iPDF;
\item the population of observed background objects differs from the
  control field population because of the effects of extinction itself
  (essentially, we will miss the faintest objects).
\end{itemize}
In the following subsections we consider all these problems in
detail.

\subsection{Intrinsic star color distribution (iPDF)}
\label{sec:intrinsic-star-color}

We assume throughout that observations are carried out on $D+1$
different magnitude bands. Not all objects need to have complete
observations in all bands.  Out of these $D+1$ magnitudes, we can
construct $D$ independent color combinations. We generally build
colors by subtracting two consecutive magnitude bands (we generalize
below to situations where one or more bands are missing).

We desire to model the stars' intrinsic color distribution, that is,
the unextinguished colors of stars measured with no error. Here we
assume that the distribution of intrinsic colors can be well described
by a Gaussian mixed model (GMM): that is, calling $\vec c$ the
$D$-dimensional vector of colors of a star, we have
\begin{equation}
  \label{eq:1}
  p(\vec{c}) = \sum_{k=1}^{K} w_k p_k(\vec{c}) \; ,
\end{equation}
where $p_k(\vec{c})$ is the distribution of the $k$-th component
($k \in \{ 1, \dots, K \}$) and $w_k$ is the associated weight. The
weights are taken to be normalized to unity, so that
\begin{equation}
  \label{eq:2}
  \sum_{k=1}^K w_k = 1 \; .
\end{equation}
Each component of the GMM is modeled as a multivariate normal
distribution with mean $\mathrm{b}_k$ and covariance $\mathcal{V}_k$:
\begin{equation}
  \label{eq:3}
  p_k(\vec{c}) = Z(\mathcal{V}_k) \exp \left[ - \frac{1}{2} (\vec{c} -
    \vec{b}_k)^\mathrm{T} \mathcal{V}_k^{-1} (\vec{c} - \vec{b}_k)
  \right] \; .
\end{equation}
The normalizing term $Z(\mathcal{V})$ takes the form
\begin{equation}
  \label{eq:4}
  Z(\mathcal{V}) = \frac{1}{\sqrt{\mathrm{det}(2 \pi \mathcal{V})}} =
  \frac{1}{\sqrt{(2 \pi)^D \, \mathrm{det} \mathcal{V}}} \; .
\end{equation}

We model the photometric error in each band using a normal
distribution, and we take the magnitude errors to be independent in
each band (as is usually the case). Calling
$\vec{m} = \{ m_1, m_2, \dots, m_{D + 1} \}$ the magnitude of a star,
the measured magnitudes are distributed as
\begin{equation}
  \label{eq:5}
  \hat{m}_i \sim \mathrm{N}(m_i, \sigma^2_i) \; .
\end{equation}
Following what we said earlier, we call the star colors
$\vec{c} = (c_1, c_2, \dots, c_D)$, where $c_i = m_i - m_{i+1}$. With
this definition, the measured colors are associated with a
correlated error, where the correlation matrix takes the form
\begin{equation}
  \label{eq:6}
  \mathcal{E} = \begin{pmatrix}
    \sigma^2_1 + \sigma^2_2 & -\sigma^2_2 & 0 & 0 & \dots & 0 \\
    -\sigma^2_2 & \sigma^2_2 + \sigma^2_3 & -\sigma^2_3 & 0 & \dots &
    0 \\ 
    0 & -\sigma^2_3 & \sigma^2_3 + \sigma^2_4 & -\sigma^2_4 & \dots &
    0 \\ 
    \vdots & \vdots & \vdots & \vdots & \ddots & \vdots \\
    0 & 0 & 0 & 0 & \dots & \sigma^2_{n} + \sigma^2_{n+1}
  \end{pmatrix} \; .
\end{equation}
Hence, a measured color $\hat{\vec{c}}$ is distributed according to
\begin{equation}
  \label{eq:7}
  p(\hat{\vec{c}}) = \sum_{k=1}^{K} w_k p_k(\hat{\mathbf{c}}) \; ,
\end{equation}
where $p_k(\hat{\mathbf{c}})$ is the distribution of the $k$-th
component over the \textup{observed }colors:
\begin{equation}
  \label{eq:8}
  p_k(\hat{\mathbf{c}}) = Z(\mathcal{V}_k + \mathcal{E}) \exp \left[ -
    \frac{1}{2} (\hat{\mathbf{c}} - \mathbf{b}_k)^\mathrm{T}
    (\mathcal{V}_k + \mathcal{E})^{-1} (\hat{\mathbf{c}} -
    \mathbf{b}_k)
  \right] \; .
\end{equation}
We note that the error matrix $\mathcal{E}$ is different for each
star, and therefore this distribution changes for each star.

We model situations where one or more bands are missing with the help
of a (rank-deficient) \emph{\textup{projection matrix}} $\mathcal{P}$:
that is, we assume that in such cases, instead of the full color
vector $\hat{\vec{c}}$, we only measure a linear combination of it
given by $\mathcal{P} \hat{\vec{c}}$. For example, for the simple case
$D = 2$, if we only measure the magnitude $m_1$ and $m_3$, that is,\
if the band 2 is missing, we set
\begin{equation}
  \label{eq:9}
  \mathcal{P} = \begin{pmatrix} 1 & 1 \end{pmatrix} \; ,
\end{equation}
so that
$\mathcal{P} \hat{\vec{c}} = \hat c_1 + \hat c_2 = \hat m_1 - \hat
m_3$. We note that $\mathcal{P} \hat{\vec c}$ also follows a Gaussian
mixture model (GMM), with means $\mathcal{P} \vec b_k$ and variances
$\mathcal{P} (\mathcal{V}_k + \mathcal{E}) \mathcal{P}^T$. With the
help of a suitable $\mathcal{P}$ matrix, we can represent all cases of
missing data, as long as at least two bands are available. We assume
that this is always the case (i.e., that data with a single band are
discarded a priori, since they are essentially useless for color
extinction measurements).

\subsection{Extreme deconvolution}
\label{sec:extr-deconv}

\begin{figure}[tb!]
  \centering
  \includegraphics[width=\hsize]{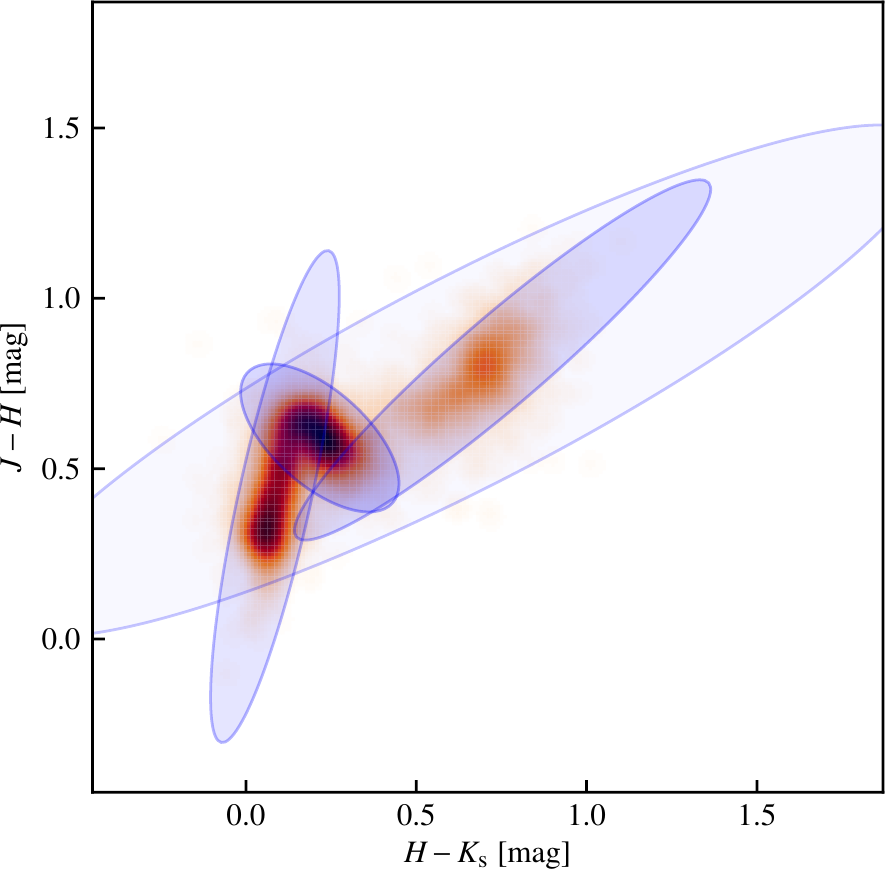}
  \caption{Extreme deconvolution of the control field
    colors. Overimposed on the same density plot as in
    Fig.~\ref{fig:1}, we show ellipses corresponding to the various
    Gaussian distributions used in the mixture that describes this
    intrinsic color probability distribution. For clarity, the
    ellipses are 50\% larger than the covariance matrices of the
    corresponding components. The ellipse fill color is proportional
    to the weight of each component. Although the model used here has
    five components, only four are evident: the fifth is a large
    ellipse that encompasses the entire figure, with a very small
    weight.}
  \label{fig:2}
\end{figure}

A critical task of XNICER is to infer the $K D (D + 3) / 2 - 1$
parameters of the GMM of the iPDF. In principle, this could be carried
out in a Bayesian framework using standard Bayesian inference, and
sampling the posterior probability with techniques such as Markov
chain Monte Carlo. In practice, this task is often non-trivial because
the control field typically includes a large number of objects
($\sim$ billions) and very many parameters.

A simpler but very effective way of solving this problem is to use a
technique called \emph{\textup{extreme deconvolution}}
\citep{2011AnApS...5.1657B}. This generalizes the well-known
expectation-maximization technique used in the K-clustering algorithm
(see, e.g., \citealp{MacKay}) to situations where the observed data
are incomplete or have errors.

The algorithm used satisfies all our requirements: it accepts noisy
and incomplete data and provides a set of best-fit parameters for a
GMM model. We note that when we use this technique, we are in the
position of using the entire control field data, including objects
with very noisy measurements, to infer the iPDF. We also note that the
results provided are, as much as possible, noise-free: that is, the
recovered parameters of the GMM model in principle only include the
intrinsic scatter in the true colors, and not the scatter caused by
their photometric errors.

We find the optimal number of components using the Bayesian
information criterion (BIC, \citealp{1978AnSta...6..461S}; see also
\citealp{2007MNRAS.377L..74L}). In practice, however, our validation
tests show that one can limit the $K$ Gaussian components to a
relatively small number ($\sim 5$--$10$) without significant
drawbacks.

Figure~\ref{fig:2} shows an example of extreme deconvolution with
$K = 5$ in the control field of the VISION dataset.

\subsection{Single-object extinction measurements}
\label{sec:single-object-extinc}

Extinction acts on the intrinsic star colors by shifting them: that
is, an extinguished star will have its colors change as
\begin{equation}
  \label{eq:10}
  \vec{c} \mapsto \vec{c} + A \vec{k} \; ,
\end{equation}
where $A$ is the extinction (in a given band) and $\vec{k}$ is the
reddening vector.

As a result, the observed distribution of an extinguished star is
\begin{equation}
  \label{eq:11}
  p(\hat{\vec{c}} \mid A) = \sum_{k=1}^K w_k p_k(\hat{\vec{c}} - A
  \vec{k}) \; ,
\end{equation}
where as before
\begin{equation}
  \label{eq:12}
  p_k(\hat{\vec{c}}) = Z(\mathcal{V}_k + \mathcal{E}) \exp \left[ -
    \frac{1}{2} (\hat{\vec{c}} - \vec{b}_k)^\mathrm{T} (\mathcal{V}_k
    + \mathcal{E})^{-1} (\hat{\vec{c}} - \vec{b}_k) \right] \; . 
\end{equation}

We now wish to derive the distribution for the extinction, that is,
$p(A \mid \hat{\vec{c}})$, using just a single star. This can be done with Bayes's theorem:
\begin{equation}
  \label{eq:13}
  p(A \mid \hat{\vec{c}}) = \frac{p(\hat{\vec{c}} \mid A)
    p(A)}{p(\hat{\vec{c}})} \; ,
\end{equation}
where $p(\hat{\vec{c}})$ is the marginalized likelihood or
\emph{\textup{evidence}}:
\begin{equation}
  \label{eq:14}
  p(\hat{\vec{c}}) = \int p(\hat{\vec{c}} \mid A') p(A') \, \mathrm{d}
  A' \; .
\end{equation}

In the simplest approach, we assume a flat prior $p(A)$ for $A$ over
the region of interest. In this way, we immediately find
\begin{equation}
  \label{eq:15}
  p(A \mid \hat{\vec{c}}) = \frac{\sum_k w_k p_k(\hat{\vec{c}} - A
    \vec{k})}{\sum_{k'} w_{k'} \int p_{k'}(\hat{\vec{c}} - A' \vec{k})
    \, \mathrm{d}A'} \; .
\end{equation}
Here the integral in the denominator, the evidence, can be used to
assess the \emph{\textup{relative}} goodness of fit of the GMM. This
can be useful to remove potential outliers in the star distribution,
that is, objects with unusual intrinsic colors (e.g., young stellar
objects, YSOs) or color measurements (e.g., spurious
detections). These objects will have likely incorrect extinction
measurements and clearly should be excluded from the analysis.

It is convenient to analyze each term in the sum in the numerator of
Eq.~\eqref{eq:15} independently. We immediately find
\begin{align}
  \label{eq:16}
  & p_k(\hat{\vec{c}} - A \vec{k}) = Z(\mathcal{V}_k + \mathcal{E})
    \notag\\ 
  & \qquad \times \exp \left[ -\frac{1}{2} (\hat{\vec{c}} - A \vec{k}
    - \vec{b}_k)^{\mathrm{T}} (\mathcal{V}_k + \mathcal{E})^{-1}
    (\hat{\vec{c}} - A \vec{k} - \vec{b}_k) \right] \; . 
\end{align}
Calling $\mathcal{W}_k = \mathcal{V}_k + \mathcal{E}$ and reorganizing
the various terms, we obtain
\begin{equation}
  \label{eq:17}
  p_k(\hat{\vec{c}} \mid A) = Z(\mathcal{W}_k) \exp \left[ -\frac{(A -
      A_k)^2}{2 \sigma^2_k} - \frac {C_k}{2} \right] \; ,
\end{equation}
where
\begin{align}
  \label{eq:18}
  & \sigma^2_k = \left( \vec{k}^T \mathcal{W}_k^{-1} \vec{k}
    \right)^{-1} \; , \\
  \label{eq:19}
  & A_k = \sigma_k^2 (\hat{\vec{c}} - \vec{b}_k)^T \mathcal{W}_k^{-1}
    \vec{k} \; , \\
  & \text{and} \notag\\
  \label{eq:20}
  & C_k = (\hat{\vec{c}} - \vec{b}_k)^{\mathrm{T}} \mathcal{W}_k^{-1}
    (\hat{\vec{c}} - \vec{b}_k) - A_k^2 \sigma_k^{-2} \; . 
\end{align}
We note that these solutions can be written more concisely if we
define a scalar product $\langle \cdot \mid \cdot \rangle$ using the
matrix $\mathcal{W}_k^{-1}$. Then we immediately have
\begin{align}
  \label{eq:21}
  & \sigma^{-2}_k = \langle \vec{k} \mid \vec{k} \rangle \; , \\
  \label{eq:22}
  & A_k = \frac{\langle \vec{c} - \vec{b}_k \mid \vec{k}
    \rangle}{\langle \vec{k} \mid \vec{k} \rangle} \; , \\
  & \text{and} \notag\\
  \label{eq:23}
  & C_k = \langle \vec{c} - \vec{b}_k \mid \vec{c} - \vec{b}_k
    \rangle - \frac{A_k^2}{\sigma_k^2} \; . 
\end{align}
This suggests that we could perform a Cholesky decomposition of
$\mathcal{W}_k = \mathcal{L} \mathcal{L}^T$ and then apply forward
substitution to calculate $\vec{v} \equiv \mathcal{L}^{-1} \vec{k}$
and $\vec{w} \equiv \mathcal{L}^{-1} (\vec{c} - \vec{b}_k)$,
quantities that can be used to compute all the rest:
\begin{align}
  \label{eq:24}
  & \sigma^{-2}_k = \vec{v}^T \vec{v} \; , \\
  \label{eq:25}
  & A_k = \sigma^2_k \vec{w}^T \vec{v} \; , \\
  & \text{and} \notag\\
  \label{eq:26}
  & C_k = \vec{w}^T \vec{w} - \frac{A_k^2}{\sigma_k^2} \; .
\end{align}
Finally, the integral appearing in the denominator of
Eq.~\eqref{eq:15}, including the component weight, is
\begin{equation}
  \label{eq:27}
  f_k \equiv w_k \int p_{k}(\hat{\vec{c}} - A' \vec{k}) \,
  \mathrm{d}A' = w_k Z(\mathcal{W}_k) \sqrt{2 \pi \sigma^2_k} \exp
  \left[ - \frac{C_k}{2} \right] \; . 
\end{equation}
Therefore, the resulting distribution for $p(A \mid \hat{\vec{c}})$ is
again a mixture of Gaussian distributions and can be written directly
as
\begin{equation}
  \label{eq:28}
  p(A \mid \hat{\vec{c}}) = \frac{\displaystyle \sum_{k=1}^K
    \frac{f_k}{\sqrt{2 \pi \sigma_k^2}} \exp \left[ - \frac{(A -
        A_k)^2}{2 \sigma_k^2} \right]}{\displaystyle \sum_{k=1}^K f_k}
  \; .
\end{equation}
The \emph{\textup{evidence}} of the measurement is provided by the
denominator. 

We note that $p(A \mid \hat{\vec{c}})$, as a function of $A$, is just
a mixture of simple univariate normal distributions. In general,
therefore, it will have up to $K$ peaks, where $K$ is the number of
components used to describe the color distribution of stars.

\subsection{Partial measurements}
\label{sec:partial-measurements}

In many cases we expect to have objects that have only partial
photometric measurements, that is,\ some missing bands. We can easily
adjust the equations written above for these cases: for this purpose,
we adopt a technique similar to the one used in the extreme
deconvolution.

As discussed above, we can introduce for each object a possibly
rank-deficient projection matrix $\mathcal{P}$, and assume that
instead of measuring the color vector $\hat{\vec c}$ of an object, we
can only measure the quantity $\mathcal{P} \hat{\vec c}$.  The
covariance matrix associated with photometric errors of
$P \hat{\vec c}$ is just $\mathcal{P} \mathcal{E} \mathcal{P}^T$. We
therefore redefine $\mathcal{W}_k$ as
\begin{equation}
  \label{eq:29}
  \mathcal{W}_k = \mathcal{P} (\mathcal{V}_k + \mathcal{E})
  \mathcal{P}^T \; .
\end{equation}
As before, we then compute the Cholesky decomposition $\mathcal{L}$ of
$\mathcal{W}_k$ and the associated vectors
$\vec{v} \equiv \mathcal{L}^{-1} \mathcal{P} \vec{k}$ and
$\vec{w} \equiv \mathcal{L}^{-1} \mathcal{P} (\vec{c} -
\vec{b}_k)$. With these quantities, we can then compute $\sigma_k$,
$A_k$, $C_k$, and $f_k$ as above in
Eqs.~(\ref{eq:24}--\ref{eq:26}).

\subsection{Mixture reduction}
\label{sec:mixture-reduction}

In some cases it might be desirable to reduce the number of components
of the extinction GMM: that is, one might wish to approximate the
mixture with a mixture with fewer components. Often, the approximation
is obtained by merging a number of components of the mixture. This
procedure is usually carried out by requiring that the merged Gaussian
preserves the first moments of the merged components. Using the
notation of the Gaussian mixture in the color space, this requires
that
\begin{align}
  \label{eq:30}
  & w_\mathrm{merged} = \sum_m w_m \; , \\
  \label{eq:31}
  & \vec{b}_\mathrm{merged} = \frac{1}{w_\mathrm{merged}} \sum_m w_m
    \vec{b}_m \; , \\
  & \text{and} \notag\\
  \label{eq:32}
  & V_\mathrm{merged} = \frac{1}{w_\mathrm{merged}} \sum_m w_m \bigl[
    V_m + (\vec{b}_\mathrm{merged} - \vec{b}_m)
    (\vec{b}_\mathrm{merged} - \vec{b}_m)^T \bigr] \; .
\end{align}
In the above equations, all sums run over the merged components
$\{ m \}$. Alternatively, one can perform pruning, that is, just
removing some components and redistributing the weights. This,
however, is generally less effective.

When applying XNICER, we typically describe the control field colors
using a handful of components. The final extinction estimates for each
object will have the same number of components as the control field
GMM. For some applications, we might wish to have a single extinction
measurement (with associated error) for each object: this is easily
achieved by using the Eqs.~(\ref{eq:30}--\ref{eq:32}) above and
merging all components. The resulting $\vec{b}_\mathrm{merged}$ will
be the extinction value associated with the object; likewise, the
resulting $V_\mathrm{merged}$ will be its variance.

\subsection{Number counts and extinction}
\label{sec:numb-counts-extinct}

\begin{figure}[tb!]
  \centering
  \includegraphics[width=\hsize]{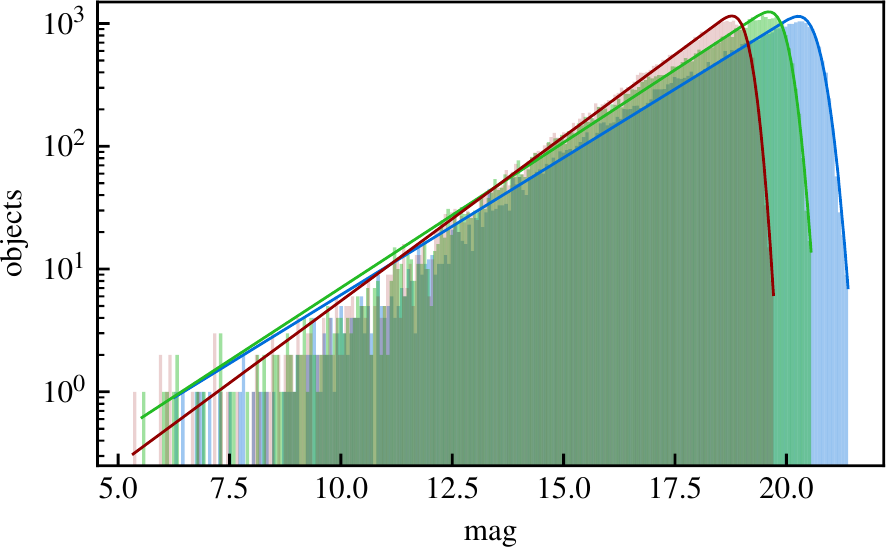}
  \caption{Histograms of the number counts in the $J$ band (blue),
    $H$ band (green), and $K_\mathrm{s}$ band (red), together with
    their best-fit models in terms of exponential distributions
    truncated by the completeness function $c(m)$ of
    Eq.~\eqref{eq:33}.}
  \label{fig:3}
\end{figure}

\begin{figure}[tb!]
  \centering
  \includegraphics[width=\hsize]{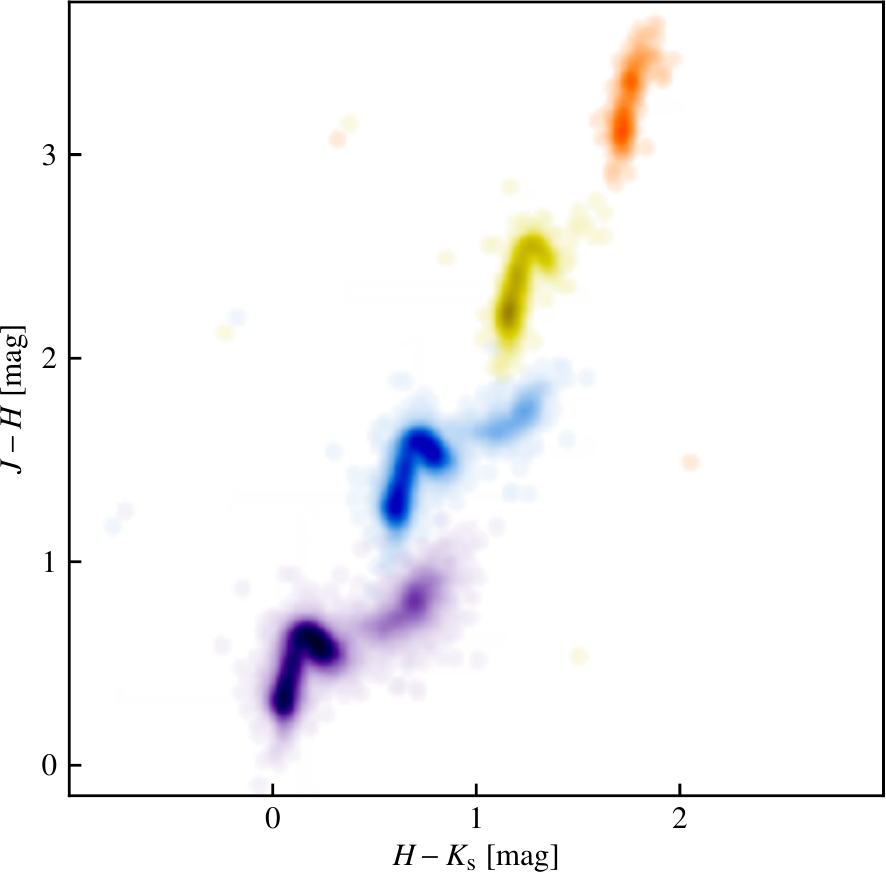}
  \caption{Effects of extinction on the color distribution, from
    $A_K = \SI{0}{mag}$ (violet) to $A_K = \SI{3}{mag}$ (orange). Not
    only is there a global shift of the distribution, but some
    components disappear progressively. The various distributions are
    renormalized: in reality, the $A_K = \SI{3}{mag}$ density is built
    from only the $3\%$ brightest objects.}
  \label{fig:4}
\end{figure}

\begin{figure}[tb!]
  \centering
  \includegraphics[width=\hsize]{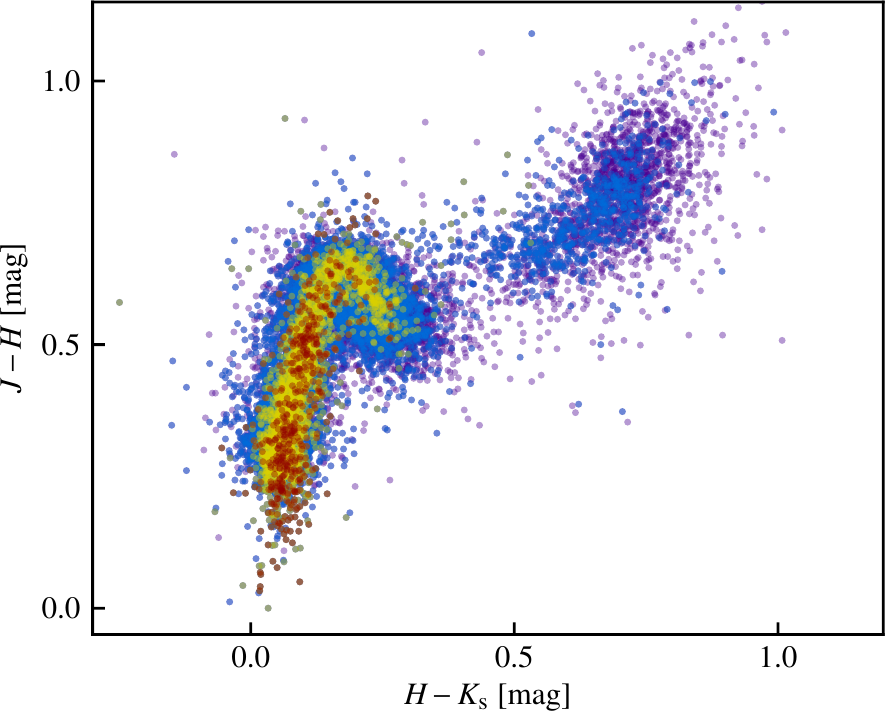}
  \caption{Change in the observed background population with
    increasing extinction, from $A_K = \SI{0}{mag}$ (violet) to
    $A_K = \SI{3}{mag}$ (red). The shift of Fig.~\ref{fig:4} is removed. The dominant effect is a change in
    relative weight of the various components, with no significant
    shift or change in shape.}
  \label{fig:5}
\end{figure}

The simple approach described above ignores a key point: because of
extinction, the population of observed stars changes, as intrinsically
faint stars will never be observed behind a dense cloud. In general,
different populations of objects will suffer in different ways from extinction: galaxies are intrinsically faint, and will be
the first objects to be wiped out by a cloud.

In order to describe the effects of extinction on the population of
objects, we need to know the \emph{\textup{completeness function}}:
this is just the probability for a star of magnitude $m$ in a given
band to be observed (at that band). In general, we will have a
completeness function in each band considered; in the case of the
VISION data considered here, we will have three functions,
corresponding to the three bands $J$, $H$, and $K_\mathrm{s}$.

We model each completeness function as a complementary error function
$\mathrm{erfc}$:
\begin{equation}
  \label{eq:33}
  c(m) = \frac{1}{2} \mathrm{erfc} \left( \frac{m - m_c}{\sqrt{2
        s^2_c}} \right) \; .
\end{equation}
In order to fit the two parameters, the 50\% completeness limit $m_c$
and the width $s_c$, we model the number counts as simple power laws.
Figure~\ref{fig:3} shows the number counts in the $J$, $H$, and
$K_\mathrm{s}$ bands measured in the VISION control field, together
with the corresponding best fits obtained for a function of the form 
\begin{equation}
  \label{eq:34}
  N(m) = N_0 \, 10^{\alpha m} \, c(m) \; ,
\end{equation}
where $N_0$ and $\alpha$ are two band-dependent constants. The fact
that the lines, representing the parametrization of Eq.~\eqref{eq:34},
are very close to the corresponding histograms shows that the selected
model for the completeness function is appropriate. This is especially
true close to the completeness limit, which is the region really
described by the three $c$ functions.\footnote{Differences in the
  bright part of the number counts might be attributed to
  different measurement techniques as well as to saturation
  effects.}\@ We stress, however, that any other functional form for
the completeness functions can be used, as XNICER is not particularly
tailored to any specific choice of $c(m)$.

Now that we know the completeness functions, that is, the two
parameters $c_m$ and $s_m$ for each band, we can simulate the presence
of a cloud and artificially remove stars that cannot be observed
because they are too faint. This process is carried out by computing
for each star the ratio $c(m + A) / c(m)$, where $A$ is the
extinction. This ratio gives for each star the probability to be
observed when there is an extinction $A$. We introduce the denominator
$1/c(m)$ in order to take into account the fact that stars in the
control field have already suffered decimation by the completeness
function.

Figure~\ref{fig:4} shows the change in population of observed
background stars when they are observed at increasing levels of
extinction. It is evident that different components have different
``survival likelihoods'': extended objects disappear almost completely
when $A = \SI{2}{mag}$, while main-sequence stars survive, although
significantly decimated, up to $A > \SI{5}{mag}$. This reflects the
different number count slopes of stars and galaxies: the former are
distributed in the infrared approximately as
\citep{2009A&A...493..735L}
\begin{equation}
  \label{eq:35}
  n(m) \propto 10^{0.33 \, m} \; ,
\end{equation}
while galaxies approximately follow Hubble's number counts (strictly
valid for a static, Euclidean universe with no galaxy evolution; see,
e.g., \citealp{Peebles}):
\begin{equation}
  \label{eq:36}
  n(m) \propto 10^{0.6 \, m} \; .
\end{equation}

The fact that different components have different behavior behind a
cloud indicates that we should use different GMMs at different
extinction levels. In order to show how the population is modified, we
plot in Fig.~\ref{fig:5} the same data as in Fig.~\ref{fig:4}, but we
also shift the points back along the reddening vector, so that the
offset in color introduced by extinction is canceled out. This plot
shows that to first order, the position and shapes of the various
components are left unaffected by extinction, and that only their
weights change (eventually vanishing if one component disappears).

We therefore adopt the following strategy to take into account the
effects of extinction in the population change. We build several GMMs
of our objects at different extinction levels. For each object, we
then use an iterative technique:
\begin{enumerate}
\item Initially, we assume for all objects a vanishing extinction and
  use the GMMs at zero extinction to infer the extinction probability
  distribution $p(A)$ for each of them.
\item We then use the $p(A)$ as a weight for the various GMMs models:
  that is, we build a GMM model that  itself is a mixture of the GMMs
  models at various extinction, weighted by the corresponding
  $p(A)$. Since the components of the various mixtures for
  different extinctions share their centers and covariances, the final
  GMM has the same number of components as the original GMM: this procedure does not add any further complexity to the
  method.
\item We repeat step 2 a few times to ensure convergence
  (which is achieved very quickly).
\end{enumerate}

\subsection{Further improvements}
\label{sec:further-improvements}

We can further improve our method by taking advantage of other
indicators that can help us to distinguish different populations of
objects. This additional step is effective if the different
populations have little superposition and if they are at different
places along the reddening vector.

One of the simplest possible strategies is using a reliable
morphological classification to separate the objects in the color
space.  In our specific case, the colors of extended objects
(galaxies) are very different from point-like objects (stars).

To proceed in this way, we perform two extreme deconvolutions for the
different classes of objects (in a way that shares some similarities
with the GNICER method of \citealp{2008ApJ...674..831F}). When
computing the extinction against one object, we then use the
corresponding class of the object.  There are two basic ways of doing
this:
\begin{itemize}
\item Hard classification: every object is taken to belong to a single
  class. In this case, we perform separate extreme deconvolutions for
  each class, and use the appropriate class for each object when
  inferring its extinction.
\item Soft classification: each object has a set of probabilities to
  belong to each class. In this case, we perform separate extreme
  deconvolutions, but using all objects with weights
  corresponding to each probability; we describe the final extinction
  probability as a mixture of the extinction probabilities we obtain
  for the various classes, weighted by the object classification
  parameter.
\end{itemize}
We opt here for the second technique, which is more general than the
first: the first technique indeed can be thought of as a particular
case of the second, where the probabilities (and thus the weights) are
either zero or one.

\section{Comparison with different techniques}
\label{sec:comp-with-diff}

In this section we briefly compare from a purely theoretical point of
view XNICER with some of the techniques that are available for
extinction studies. A discussion of the various performances is
carried out in the next section.

\subsection{NICER}
\label{sec:nicer}

As noted above, XNICER generalizes NICER, in the sense that it
decomposes the iPDF in terms of a mixture of Gaussian distributions,
while NICER only uses (implicitly) a single Gaussian. In order to
better understand and quantify this statement, however, it is useful
to reformulate NICER from a perspective that is similar to XNICER.

NICER is implemented by considering the distribution of star colors in
the control field: there, for objects with accurate photometric
measurements in all bands, one measures the average color and
associated covariance matrix. In the limit of negligible photometric
errors, this is totally equivalent to an extreme deconvolution with a
single component.

NICER then combines the control field calibration with the colors of
each object to infer the extinction. The corresponding equations are
virtually identical to Eqs.~\eqref{eq:19} and \eqref{eq:18} for the
extinction estimate and associated error, respectively. That is,
again, NICER is a special case of a single component XNICER.

In spite of these similarities, the NICER implementation shows some
significant differences if one looks in detail at its
implementation. The fact that NICER does not take directly into
account the photometric errors in the control field generally result
in an inaccurate calibration, as we show below. This has two
important consequences: extinction measurements can be (slightly)
biased, and the computed errors are overestimated. We quantify
both these effects for the VISION data below.

A second important difference is the lack in NICER of a correction for
the population change for increasing extinction. This again results in
a bias in the extinction estimated (which in general increases with
extinction).

Additionally, it is worth noting that the amount of the two
inaccuracies we described depends on the depth of the data used. NICER
has been developed and generally used with relatively shallow data,
such as are offered by the Two Micron All Sky Survey
(2MASS). Additionally, the use of NICER has generally been restricted
to point-like objects. Together, all this limits the biases present in
NICER, since the galaxy component in the color-color plot (which is
responsible for most of the population bias, the most severe one) is
absent.

\subsection{SCEX}
\label{sec:scex}

\citet{2016A&A...585A..78J} have developed SCEX, a series of
relatively complex techniques that also aim at representing the
distribution of the intrinsic colors more precisely than NICER. SCEX,
in contrast to XNICER, adopts a fully non-parametric approach to
describe the iPDF, which is directly inferred using kernel density
estimation (KDE) techniques. Thus SCEX might be thought to be better
at capturing peculiarities in the iPDF, since KDE surely allows more
freedom than GMM (unless an exceedingly larger number of components is
used, which generally is not the case).

In reality, precisely because of the use of KDE techniques, SCEX is
completely unable to cope with photometric errors, both in the control
field and in the science field. The control field color PDF is
obtained from a smoothing of the star colors with a fixed kernel of
\SI{0.1}{mag}. However, this procedure does not lead to an estimate of
the iPDF, but rather provides something that is close to a smoothed
distribution of \textit{\textup{observed}} colors. Additionally,
photometric errors are used in a very simplified way for the
extinction estimate.

\subsection{PNICER}
\label{sec:pnicer}

PNICER uses some of the ideas of SCEX and therefore shares several
similarities with XNICER. It also adopts a fully non-parametric
approach to describe the iPDF, which is directly inferred using KDE
techniques. Therefore, similarly to SCEX, PNICER is in principle able
to better describe complex star color distributions than
XNICER. Additionally, when used on single objects, instead of a simple
extinction measurement with associated error, it can return a
representation of the probability distribution of the extinction in
terms of Gaussian mixtures. XNICER does exactly the same in a direct
and natural way.

In order to understand the limitations of PNICER, it is useful to
consider its implementation in detail. This algorithm works by
building a representation of the control field either in magnitude or
color space. Essentially, it takes all control field objects that have
complete measurements on a given set of passbands and computes the
density distribution using a KDE based on simple kernels with fixed
bandwidth. The user is free to perform the calculations in magnitude
space or color space. For a color-based PNICER, the probability that a
star has a given color combination is obtained by counting how many
stars in the control field have similar colors; something similar, but
in magnitude space, happens for a magnitude-based PNICER. This
approach necessarily introduces a number of simplifications:
\begin{itemize}
\item First, similarly to SCEX, the KDE technique itself forces the
  code to use a single scale for the kernel. In particular, the kernel
  is the same not only for each star (i.e., stars with wildly
  different photometric errors are still used in the same way for the
  density estimate), but also for different color combinations. In its
  current implementation, the bandwidth of the kernel is chosen by
  taking the mean error along all color combinations.
\item PNICER, in contrast to XNICER, does not try in any way to obtain
  the intrinsic color (or magnitude) probability distribution iPDF: it
  only works on the observed distribution.
\item Additionally, when inferring the extinction of a star, PNICER
  ignores all star photometric errors. Simply, the observed star
  colors are taken as the true ones, and the error on the star
  extinction is just the result of the width of the control field
  color distribution along the reddening vector originating from the
  star.
\item Finally, in its current implementation, PNICER ignores all
  selection effects introduced by extinction; these are discussed in
  Sect.~\ref{sec:numb-counts-extinct}.
\end{itemize}
The lack of an appropriate error treatment in PNICER also implies that
the errors estimated by this method are inaccurate, and indeed, our
tests have shown that PNICER generally overestimates the true
extinction errors by almost a factor of 2.

\section{Control field analysis}
\label{sec:contr-field-analys}

As a simple test, we used the same VISION control field stars to check the performance of XNICER. For our analysis we took the control field photometric measurements and split them randomly into
two sets of the same size. We then considered one of these two
sets as a control field and used it to train the various algorithms,
and the other as a sort of science field. In all cases, we asked
the algorithms to return a single estimate for each object: this is
the only possibility of NICER, and also the standard output returned
by PNICER. XNICER, instead, by default provides a GMM of the inferred
extinction for each object: for a more direct comparison with the
other techniques, the returned GMM was converted into a single
Gaussian using Eqs.~(\ref{eq:30}--\ref{eq:32}).

For our tests, we considered two cases, considered separately in
the following subsections:
\begin{itemize}
\item Absence of any extinction, so that the ``science field'' is used
  without any modification.
\item Presence of a constant extinction value. In this case, we
  added to the magnitudes of all objects in the science field a
  suitable value (which depends on the extinction law) and
  simulated the effects of incompleteness.
\end{itemize}

In order to quantify the merits and weaknesses of each algorithm, it
is necessary to define some quantities that summarize the bias and the
average noise. Specifically, calling $A_n$ and $\sigma_n$ the
extinction measurement and its associate error for the $n$-th star, we
define
\begin{align}
  \label{eq:37}
  b & {} \equiv \frac{\displaystyle \sum_{n=1}^N
  \frac{A_n - A_\mathrm{true}}{\sigma^2_n}}{\displaystyle \sum_{n=1}^N
      \frac{1}{\sigma^2_n}} \; , \qquad &
  e & {} \equiv \frac{\displaystyle \sqrt{N \sum_{n=1}^N
      \left(\frac{A_n - A_\mathrm{true}}{\sigma_n^2} \right)^2}}{\displaystyle
      \sum_{n=1}^N \frac{1}{\sigma_n^2}} \; ,
\end{align}
where $A_\mathrm{true}$ is the true extinction value (set to zero in
the next subsection, and to \SI{1}{mag} or \SI{2}{mag} in the
following one). The use of the $1/\sigma^2_n$ weights in the above
equations is justified by simple statistical arguments: if the
$\{ \sigma_n \}$ are reliable estimates of the measurement errors
for the $\{ A_n \}$, the chosen weights minimize the final
variance. For this reason, these weight terms are used very
often when different extinction measurements need to be averaged
together, as in many extinction maps of molecular clouds (see, e.g.,
\citealp{2010A&A...512A..67L, 2014A&A...565A..18A,
  2018arXiv180301004M}).

In summary, $b$ can be interpreted as the bias, and $e$ can be
interpreted as the average total error associated with each extinction
measurement. In the way it is defined, $e$ includes both the
statistical error (associated with the scatter of the various $A_n$)
and the systematic error (associated with the bias $b$).

\textit{\textup{In the following we mainly use $e$ as a measurement of
    the expected total error of each method}.} Ideally, we would like
to have $e$ as small as possible and $b$ as close as possible to
zero. Additionally, it would be desirable that the average estimated
error, $\langle \sigma_n \rangle$, be close to $e$, indicating that
the error estimate is reliable. An inspection of the definition of $e$
in Eq.~\eqref{eq:37} also shows that the quantity $e$ reduces to
\begin{equation}
  \label{eq:38}
  \bar e \equiv \sqrt{\frac{N}{\sum_{n=1}^N 1/\sigma_n^2}} \; 
\end{equation}
if $\sigma_n$ is correctly estimated and the bias $b$
vanishes. Therefore, in the following we also consider $\bar e$ to
verify that the various algorithms correctly estimate their errors, so
that $e \simeq \bar e$.

\subsection{No extinction}
\label{sec:no-extinction}

\begin{figure}[tb!]
  \centering
  \includegraphics[width=\hsize]{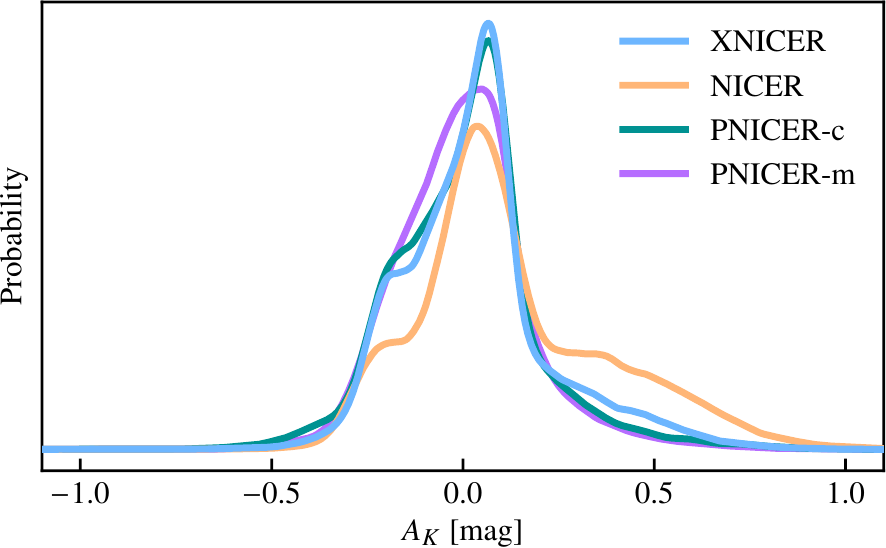}
  \caption{The distribution of extinction for the control field stars,
    as determined using XNICER, NICER, and PNICER, both in color space
    (PNICER-c) and magnitude space (PNICER-m), shown using a KDE
    with size \SI{0.05}{mag}.}
  \label{fig:6}
\end{figure}

\begin{figure}[tb!]
  \centering
  \includegraphics[width=\hsize]{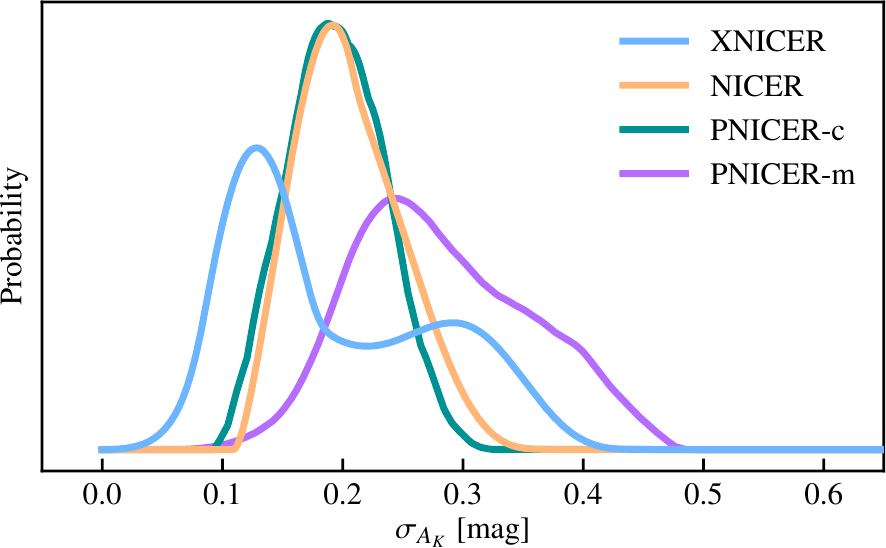}
  \caption{The distribution of nominal errors for the extinction
    values for the control field stars, as determined using XNICER,
    NICER, and the two PNICER versions.}
  \label{fig:7}
\end{figure}

\begin{figure}[tb!]
  \centering
  \includegraphics[width=\hsize]{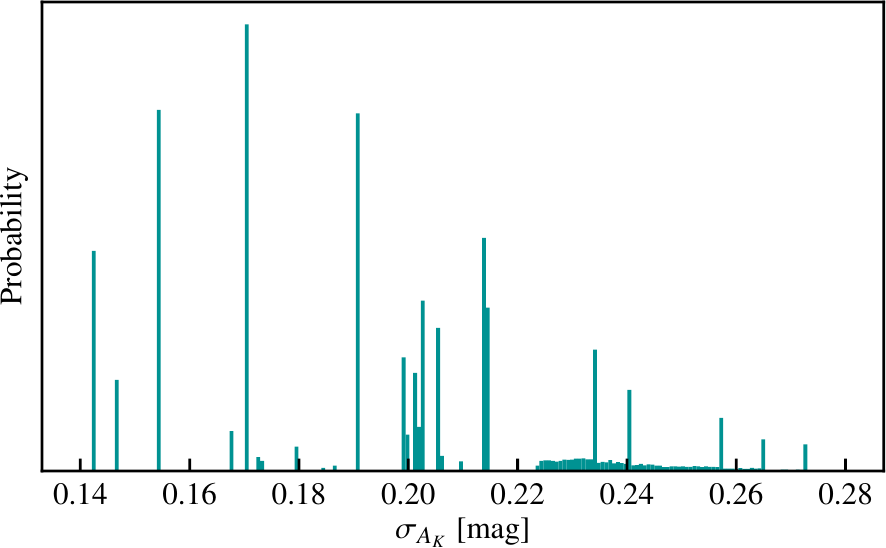}
  \caption{Distribution of nominal errors PNICER-c, shown as
    a histogram. Many objects show the same nominal error.}
  \label{fig:8}
\end{figure}

\begin{table}[tb!]
  \caption{Control field results. Each table cell shows the
    combination $b \pm e$, that is, the bias and the average
    error associated with a single extinction measurement computed
    using Eq.~\eqref{eq:37} for different true extinction values
    (columns) and different methods (rows). These values are followed
    by the expected noise $\bar e$, evaluated using
    Eq.~\eqref{eq:38}. Ideally, $b$ should be as close to
    zero as possible (negligible bias), $e$ should be as small as
    possible (small noise), and $\bar e$ should be as close as
    possible to $e$ (accurate noise estimate). All quantities are
    expressed in magnitudes.}
  \label{tab:1}
  \scriptsize
  \begin{tabular}{lccc}
    \hline\hline
    Method & $A_K = \SI{0}{mag}$ & $A_K = \SI{1}{mag}$ & $A_K = \SI{2}{mag}$ \\
    \hline
    XNICER   & $\phantom{-}0.00 \pm 0.14$ $[0.14]$ 
             & $-0.01 \pm 0.12$ $[0.11]$
             & $-0.03 \pm 0.11$ $[0.10]$\\
    NICER    & $+0.07 \pm 0.22$ $[0.20]$ 
             & $-0.03 \pm 0.15$ $[0.17]$
             & $-0.11 \pm 0.19$ $[0.16]$\\
    PNICER-c & $\phantom{-}0.00 \pm 0.18$ $[0.19]$
             & $-0.04 \pm 0.13$ $[0.16]$
             & $-0.08 \pm 0.14$ $[0.16]$\\
    PNICER-m & $\phantom{-}0.00 \pm 0.15$ $[0.26]$
             & $-0.01 \pm 0.26$ $[0.16]$
             & $-0.01 \pm 0.57$ $[0.01]$\\
    \hline
  \end{tabular}
\end{table}

In the first version of the check, we used the ``science field'' data
without any modification. As a result, we expect to observe a
vanishing extinction on average.

Figure~\ref{fig:6} shows the extinction distributions\footnote{We
  produced this figure, as well as many of the following ones, using a
  KDE technique with kernel size equal to \SI{0.05}{mag}.} obtained in
the control field using XNICER, NICER, and the two PNICER variants,
which we call PNICER-c (for the color-based one) and PNICER-m (for the
magnitude-based one).\footnote{For all tests we used the latest PNICER
  code available at \url{https://github.com/smeingast/PNICER}. At the
  time of our tests, the latest PNICER version was v0.1-beta1.}\@ For
this and for the entire following analysis, we made use of the
\citet{2005ApJ...619..931I} extinction law. It is already evident from
this figure that XNICER and the two PNICER versions show significantly
narrower distributions than NICER. This qualitative statement is
confirmed by the first column of Table~\ref{tab:1}. Moreover, PNICER-m
performs better than PNICER-c, which is not immediately evident from
Fig.~\ref{fig:6}: presumably, this is related to a better estimate of
the weights by PNICER-m. XNICER and the two PNICER versions also have
a negligible bias $b$.

Figure~\ref{fig:7} shows the distribution of noise estimates for the
various methods. Interestingly, XNICER has a relatively wide
distribution: in practice, it strongly distinguishes between different
objects, and the noise estimate spans one order of magnitude. This
distribution has a peak at very low values of $\sigma_{A_K}$ , which
also implies that XNICER will give relatively high weights to some
objects. In contrast, PNICER-c and NICER have narrow (and very
similar) noise distributions, and will therefore use little weighting
in the Eq.~\eqref{eq:37}. PNICER-m, instead, has a large noise
distribution, shifted toward relatively high values. This wide
distribution is the result of an inaccurate noise estimate: as
indicated in Table~\ref{tab:1}, PNICER-m predicts a noise level of
$\bar e = \SI{0.26}{mag}$, but the real noise is only
$\SI{0.15}{mag}$.

Judging from both Fig.~\ref{fig:7} and Table~\ref{tab:1}, PNICER-c
seems to have more consistent noise properties. In reality, the
algorithm shows some peculiarities in its current version, highlighted
by Fig.~\ref{fig:8}: many objects are reported to have the same formal
noise. This is the result of some binning operated in the PNICER-c
implementation, which is necessary to implement a high speed of the
algorithm. Although this does not seem to cause any severe problem for
the use of the algorithm, we speculate that it might make the
algorithm less efficient and might be related to the fact that
PNICER-c has a slightly higher noise than PNICER-m.

\subsection{Non-vanishing extinction}
\label{sec:non-vanish-extinct}

\begin{figure}[tb!]
  \centering
  \includegraphics[width=\hsize]{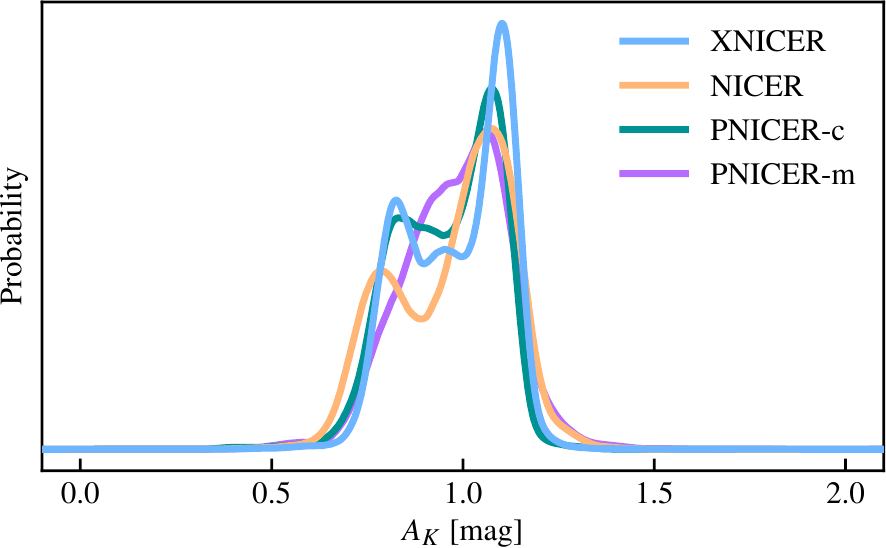}
  \caption{Distribution of extinction measurements for a star of
    the control field when an extinction of $A_K = \SI{1}{mag}$ is
    artificially applied.}
  \label{fig:9}
\end{figure}

\begin{figure}[tb!]
  \centering
  \includegraphics[width=\hsize]{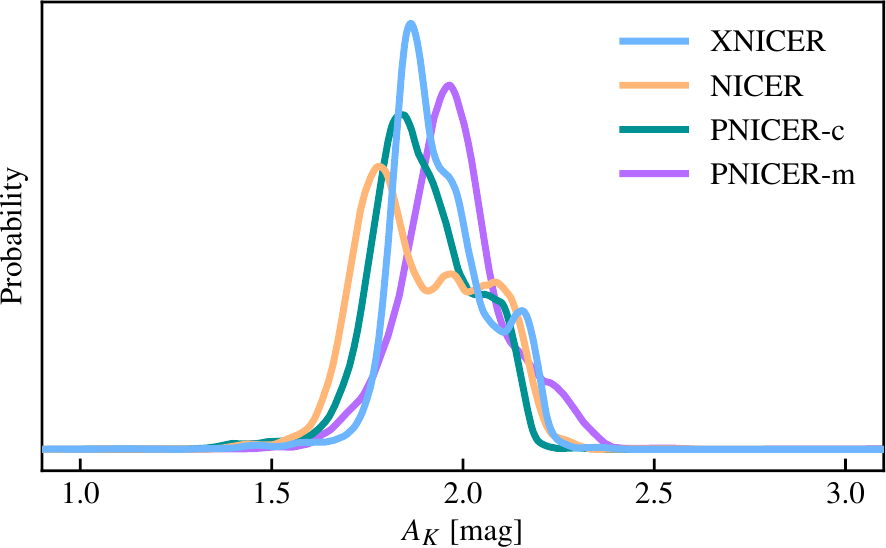}
  \caption{Same as Fig.~\ref{fig:9}, but for an extinction of $A_K =
    \SI{2}{mag}$.}
  \label{fig:10}
\end{figure}

We can easily simulate the effects of extinction by adding a term proportional to the reddening vector to each band. We then simulate
the effects of incompleteness by statistically dropping photometric
measurements for faint stars, similarly to what was done in
Sect.~\ref{sec:numb-counts-extinct}.

Figures~\ref{fig:9} and \ref{fig:10} show the results obtained for the
various methods for a $A_K = \SI{1}{mag}$ and a $A_K = \SI{2}{mag}$
cloud. The corresponding $b$, $e$, and $\bar e$ values are shown in
Table~\ref{tab:1}. Several simple conclusions can be carried out.

First, we are reassured that XNICER is able to cope well with both
moderate and high extinction values: the bias is always very limited,
and the error even decreases. This latter effect arises because as extinction increases, we miss the galaxy blob, and
therefore the scatter in the intrinsic color of objects
decreases. The formal error, $\bar e$, is also always very
close to the actual error $e$.

Second, it is interesting to note that NICER retains much of its power
even at relatively high extinction values. Admittedly, it suffers from
some bias; however, it is still confined within \SI{0.1}{mag} at
$A_K = \SI{2}{mag}$, while its noise never increases. All this is most
likely due to the simple algorithm that is used, which is able to cope
well with the change of the iPDF introduced by the extinction.

Third, PNICER-c performs well as $A_K$ increases: it does not match
the precision of XNICER, but is not far from it. It suffers some bias,
which increases with extinction, but is still below \SI{0.1}{mag}.

Instead, even a moderate extinction is able to severely affect
PNICER-m. This is related to its inaccurate noise estimate: the values
of $\bar e$ decrease from \SI{0.26}{mag} for $A_K = \SI{0}{mag}$ (a
value that, as we noted, is overestimated) to \SI{0.01}{mag} for
$A_K = \SI{2}{mag}$ (a value that is clearly underestimated).

Finally, we report the result of a further test, where we apply the
zero-extinction XNICER deconvolution to the extinguished control field
stars. In this case, we measure a moderate bias, similar to the bias
of PNICER-c (\SI{0.03}{mag} for $A_K = \SI{1}{mag}$ and \SI{0.07}{mag}
for $A_K = \SI{2}{mag}$), together with a very small increase in the
noise ($e = \SI{0.13}{mag}$ for $A_K = \SI{2}{mag}$ and no increase at
lower extinctions). These results indicate that only a small penalty
is due if one desires to use a simpler algorithm and avoid the
multiple fits at different extinctions.

\section{Sample application: Orion~A}
\label{sec:sample-appl-orion}

\begin{figure}[tb!]
  \centering
  \includegraphics[width=\hsize]{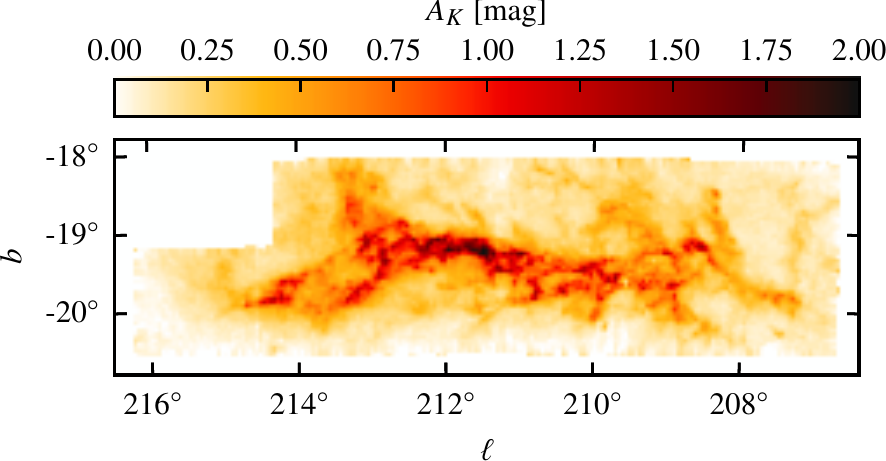}
  \caption{XNICER extinction map of Orion~A, using VISION data.}
  \label{fig:11}
\end{figure}

\begin{figure}[tb!]
  \centering
  \includegraphics[width=\hsize]{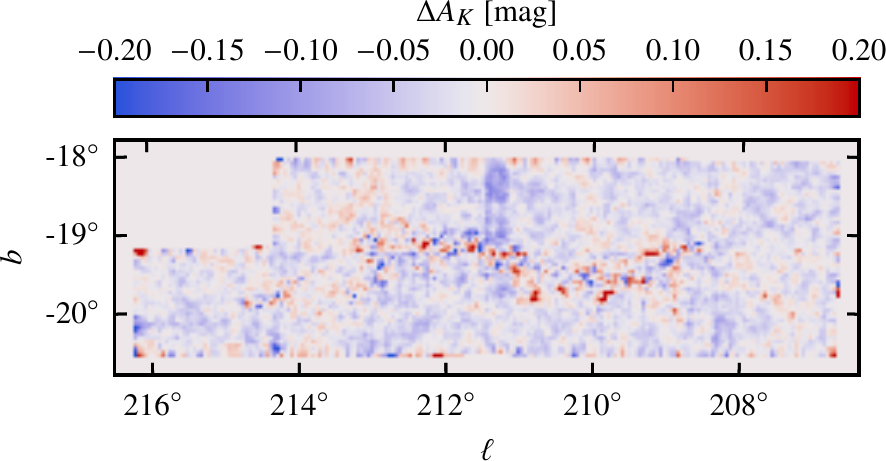}
  \caption{Differences between the XNICER and the NICER extinction map
    of Orion~A, defined as
    $\Delta A_K = A_K^\mathrm{(XNICER)} - A_K^\mathrm{(NICER)}$.}
  \label{fig:12}
\end{figure}

\begin{figure}[tb!]
  \centering
  \includegraphics[width=\hsize]{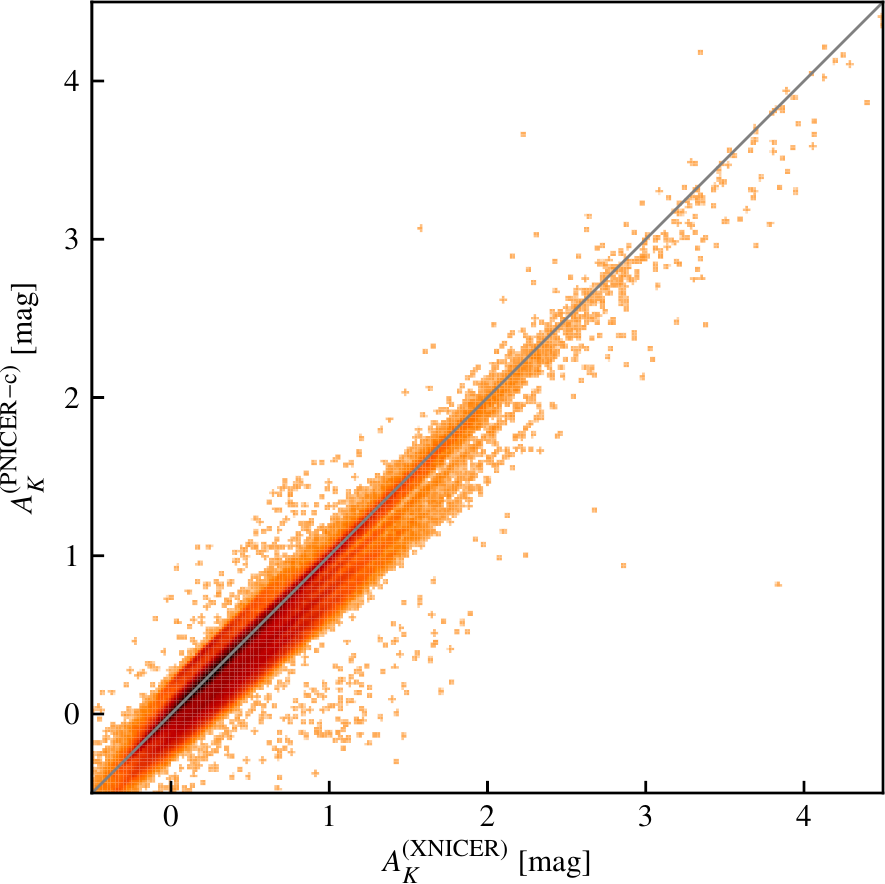}
  \caption{Object-to-object comparison of the XNICER and PNICER-c
    extinction measurements in the Orion~A field.}
  \label{fig:13}
\end{figure}

\begin{figure}[tb!]
  \centering
  \includegraphics[width=\hsize]{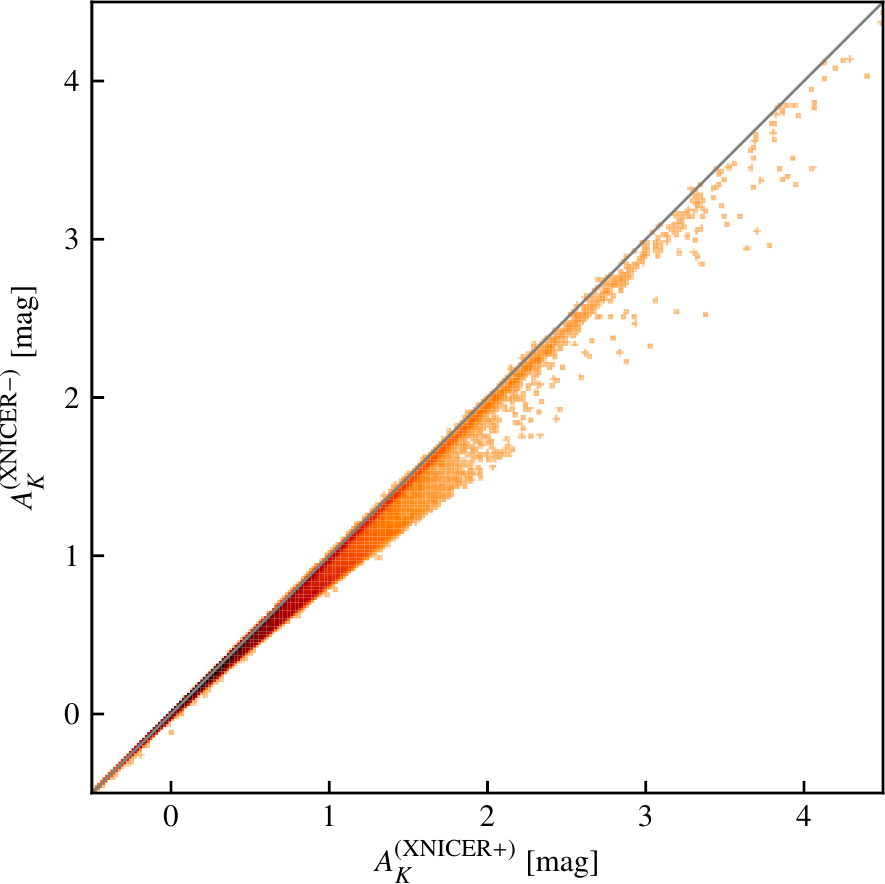}
  \caption{Comparison between the XNICER extinction estimates obtained
    with the effects of the incompleteness (XNICER+) and without
    them (XNICER-).  XNICER- can underestimate the
    extinction.}
  \label{fig:14}
\end{figure}

As a first case, we applied our method to the Orion~A VISION data. For
this purpose, we computed the extinction against each object using
NICER, PNICER-c, and XNICER. We then averaged all measurements using a
standard moving-average technique, as is typically employed for
extinction studies: calling $A_n$ and $\sigma_n$ the XNICER extinction
and estimated noise for the $n$-th object, we computed the extinction
at the sky position $\vec x$ as
\begin{equation}
  \label{eq:39}
  A(\vec x) = \frac{\displaystyle \sum_n W(\vec x - \vec x_n)
    \frac{A_n}{\sigma^2_n}}{\displaystyle \sum_n \frac{1}{\sigma^2_n}}
  \; .
\end{equation}
Here $\vec x_n$ is the angular position of the $n$-th object, and $W$
is a suitable weight function, taken to be normalized to unity:
\begin{equation}
  \label{eq:40}
  \int W(\vec x') \, \mathrm{d}^2 x' = 1 \; .  
\end{equation}
In this sample application, we used for $W$ a two-dimensional Gaussian
with $\mathit{FWHM} = \SI{2.4}{arcmin}$.  As before, we trained XNICER
using a five-component Gaussian mixture, but then we merged each
extinction measurement into a single value (with associated noise) by
joining the various components. More sophisticated techniques for
extinction spatial averaging will be considered in a follow-up
paper. The result obtained, reported in Fig.~\ref{fig:11}, shows that
there are no obvious problems with the algorithm.

Figure~\ref{fig:12} shows the difference between the XNICER and the
NICER extinction maps. This figure is dominated by relatively small
fluctuations that are due to the different noise properties of the two
algorithms. An investigation of the predicted noise maps shows indeed
that XNICER provides overall extinction measurements in the field
whose noise is a factor $\sim 2$ lower, consistent with the data shown
in Table~\ref{tab:1}. The most relevant differences are observed in
the densest regions, where XNICER generally provides higher extinction
values (red dots in the figure): again, this is consistent with the
results shown in Table~\ref{tab:1}, where we see that NICER generally
underestimates the extinction for relatively high extinction
values. Finally, the blue band that is observable in the top center of
Fig.~\ref{fig:12} is most likely the result of some systematic effects
present in the VISION data.
 
As an additional test, we compared object-by-object the extinction
estimated using XNICER with the extinction estimated using NICER and
PNICER-c; we did not use PNICER-m here because this technique, as
discussed in Sect.~\ref{sec:contr-field-analys}, does not sample the
control field for the VISION data well enough.

The results of a comparison between XNICER and PNICER-c are shown in
Fig.~\ref{fig:13}. In general, we see a good match between the
extinction estimates. Some of the scatter can be simply attributed to
the differences in the algorithms. However, we also see a trend at
high column densities: XNICER tends to return slightly higher
extinction values than PNICER-c. This is consistent with our finding
and with the results shown in Table~\ref{tab:1}, where we see that
PNICER-c can underestimate the extinction because it ignores the
effects of incompleteness.

We further checked this point by comparing the XNICER extinction
estimates obtained with and without incompleteness correction. The
result, shown in Fig.~\ref{fig:14}, confirms the trend of
Fig.~\ref{fig:13}: if the incompleteness is not taken into account, we
underestimate the extinction.

\section{Implementation}
\label{sec:implementation}

XNICER has been implemented in Python. It uses standard scientific
libraries (\texttt{numpy}, \texttt{scipy}, and
\texttt{scikit-learn}). Additionally, it relies on the extreme
deconvolution algorithm\footnote{See
  \url{https://github.com/jobovy/extreme-deconvolution}} of Bovy et
al., which is available as a dynamic C library with (among others)
Python wrapper. The library can take advantage of parallel processing
capabilities through the OpenMP programming interface.

The extreme deconvolution of the control field data, which can be
considered as a training phase in the algorithm, is a critical step
and usually the most computationally intensive one. The execution time
heavily depends on the number of components that are fit: however,
during our tests, the training on the VISION data (comprising slightly
more than $80\,000$ objects) has required approximately one minute for
five components on a personal computer. We therefore expect that
extreme deconvolutions on much larger datasets should be possible
without difficulties using more powerful workstations.

The rest of the algorithm does not pose any particular computational
challenge and is executed in a fraction of a second for large
datasets. This is advantageous since in normal application, the
science field contains many more stars than the control field where
the training step is carried out.

In summary, we do not envisage any particular technical difficulty to
apply XNICER to large datasets. We plan to release the full code
shortly after the publication of this paper.

\section{Conclusions}
\label{sec:conclusions}

We have presented a new method for performing precise
extinction measurements using arbitrary multi-band observations. The
method, implemented in the Python language, enjoys a number of useful
properties that we summarize in the following points:
\begin{itemize}
\item It provides a reliable description of the intrinsic colors of
  stars in terms of a Gaussian mixture model. For this purpose, it
  makes use of the extreme deconvolution technique.
\item It is based on rigorous Bayesian statistical arguments and fully
  takes into account all noise properties of the control field and
  science field objects. Moreover, it can be applied without any
  difficulty to objects with partial photometric measurements
  (measurements only in a subset of the available bands).
\item It provides the extinction probability distribution directly for
  each object. This quantity is returned already decomposed in terms
  of a GMM. If required by the specific application, the full output
  can be reduced to two simple quantities: the object mean extinction,
  and its associated error.
\item The method has been further improved to take into account the
  effects of incompleteness as a consequence of an increasing
  extinction. Furthermore, it can also make use of additional
  information, such as the object morphology.
\item Our tests have shown that the method performs very well under a
  range of conditions and outperforms the NICER and PNICER
  techniques. In particular, we have shown using the Orion~A VISION
  data that XNICER reduces the noise of extinction measurements by
  approximately a factor two with respect to the NICER algorithm, and
  that, in contrast to NICER and PNICER, it is not affected by any
  significant bias even at high extinctions.
\end{itemize}

\begin{acknowledgements}
  We thank the anonymous referee for helping us to improve the paper
  significantly. We are grateful to Stefan Meingast and Jo\~ao Alves
  for their constructive comments on this work. The code at the base
  of this research has made use of Astropy, a community-developed core
  Python package for Astronomy \citep{2013A&A...558A..33A}.
\end{acknowledgements}

\bibliographystyle{aa} 
\bibliography{dark-refs}

\end{document}